\documentclass[aps,superscriptaddress,twocolumn,showkeys]{revtex4-1}

\usepackage{amsmath}
\usepackage{float}
\usepackage{amsfonts}
\usepackage{graphicx}
\usepackage{bmpsize}
\usepackage{xcolor}
\usepackage{epstopdf}
\usepackage{xspace}

\newcommand{\UVfalse}{\ensuremath{(U,V)_{\false}}\xspace}
\newcommand{\UVtrue}{\ensuremath{(U,V)_{\true}}\xspace} 
\newcommand{\true}{\textsc{true}\xspace}
\newcommand{\false}{\textsc{false}\xspace}

\newcommand{\ie}{i.\,e.\xspace}
 
\newcommand{\eg}{e.\,g.\xspace}

\newcommand{\deriv}[1]{\ensuremath{\frac{{\rm d}#1}{{\rm d}t}}}

\begin{document}


\title{Dynamic modulation of external conditions can transform chemistry into logic gates}



\author{Matthew Egbert}
\email[]{m.egbert@auckland.ac.nz}
\affiliation{Department of Earth and Planetary Sciences, Harvard University, Cambridge, Massachusetts, USA}
\affiliation{University of Auckland, Auckland, NZ}

\author{Jean-S\'{e}bastien Gagnon}
\email[]{gagnon01@fas.harvard.edu}
\affiliation{Department of Earth and Planetary Sciences, Harvard University, Cambridge, Massachusetts, USA}

\author{Juan P\'{e}rez-Mercader}
\email[]{jperezmercader@fas.harvard.edu}
\affiliation{Department of Earth and Planetary Sciences, Harvard University, Cambridge, Massachusetts, USA}
\affiliation{Santa Fe Institute, Santa Fe, New Mexico, USA}


\date{\today}

\begin{abstract}
We introduce a new method for transforming chemical systems into desired logical operators (\eg NAND-gates) or similar signal-manipulation components. The method is based upon open-loop dynamic regulation, where external conditions such as feed-rate, lighting conditions, etc. are modulated according to a prescribed temporal sequence that is independent of the input to the network.  The method is first introduced using a simple didactic model. We then show its application in transforming a well-stirred cubic autocatalytic reaction (often referred to as the Selkov-Gray-Scott model) into a logical NAND-gate.  We also comment on the applicability of the method to biological and other systems.
\end{abstract}


\maketitle


\section{Introduction \label{sec:Introduction}}

Natural computing is a broad field of research at the intersection of
computer science, physics, chemistry and biology.  Its two main areas are the study
of biology-inspired computing (\eg neural networks and genetic algorithms), and
information processing in organisms, including for example the computational aspects
of self-assembly and biochemical reactions ---see Ref.~\cite{Rozenberg_etal_2012} for
a review.  These research areas are largely inspired by the impressive adaptivity
demonstrated by living systems. Even the simplest single-celled organisms are capable
of adaptation, and whether one sees this as a computational process (\eg~
\cite{Bray_1995,Mitchell_2011}), or one focuses upon sensorimotor feedback dynamics
\cite{Beer_1997,van_Gelder_1998}, it is interesting to consider how these
organisms, which lack nervous tissues, can use chemical processes to `process
information,' \ie to transform sensory input into functional motor output.

A recurring goal in this area is to implement chemistry-based `information processing
units' such as logic-gates. Efforts to build logic-gates and other signal-processing
devices out of chemistry date back at least as far as 1974~\cite{Rossler_1974}, and a
range of approaches, both theoretical and experimental, have been taken. These
involve diverse chemical media, including biochemical systems (\eg, enzyme reaction
networks~\cite{Okamoto_etal_1987,Okamoto_1992}, gene-regulatory networks
\cite{Hasty_etal_2002,Wang_etal_2011,Bradley_etal_2016}, peptide-based computing~\cite{Ashkenasy_etal_2011} and DNA-computing \cite{Paun_etal_2005}); abiotic chemistry (\eg Belousov-Zhabotinsky
oscillators ~\cite{Wang_etal_2016}, the bistable iodate-arsenous acid
reaction~\cite{Laplante_etal_1995}, or other molecules~\cite{DeSilva_Uchiyama_2007}); and mathematical idealizations or abstractions
of chemistry (sometimes referred to as `artificial chemistry'
\cite{Dittrich_etal_2001}), where networks are designed to satisfy various
desiderata and the constraints of actual chemistry are, for the moment, ignored. One
example of this last class of study can be found in the work of Hjemfelt et al.
\cite{Hjelmfelt_etal_1991,Hjelmfelt_etal_1992a,Hjelmfelt_etal_1992b}, who proposed a
reaction-motif that recreates McCulloch-Pitts neuron dynamics and that can be
assembled into networks that accomplish Boolean operations. This work was then used
to implement a chemistry based pattern-generator neural network using the bistable
iodate-arsenous acid reaction run in a continuous flow stirred-tank reactor (CSTR)~\cite{Laplante_etal_1995}.

The target behaviours in these studies include switching and rectification
\cite{Okamoto_etal_1987}, mimicking the behaviour of neurons
\cite{Hjelmfelt_etal_1991,Hjelmfelt_etal_1992a,Hjelmfelt_etal_1992b}, and other
functions, but perhaps the most popular goal has been to recreate the Boolean
operations of logic gates such as AND, XOR, NAND etc. Part of the appeal of
logic-gates is that they are sufficient for implementing flip-flop memory units,
half-adders and the other networks from which modern computers are made.

Often, the first step for implementing chemistry-based logic-gates is to decide how
the Boolean values (\true and \false) will be represented. The most intuitive way to
do so is to equate certain reactant concentrations with Boolean values (\eg high
acidity = \true; low acidity = \false) and to identify or develop chemical reactions
that transform those concentrations in ways that correspond to the desired Boolean
operations. Other representations are also possible. For instance in excitable
media~\cite{Toth_Showalter_1995}, the presence or absence of chemical waves can be
used to play the role of binary logic values, chemical wave propagation in narrow
capillary tubes play the role of information transmission between inputs and outputs,
and the specific geometric configuration of capillary tubes dictates the computation
to be done.  A similar idea is explored in Ref.~\cite{Steinbock_etal_1996}, where the
network of capillary tubes is replaced by a `printed circuit' of catalyst.  This idea
of using excitable media is pushed further by Adamatzky and
co-workers~\cite{Adamatzky_2004,Costello_Adamatzky_2005} (see
also~\cite{Adamatzky_etal_2005,Adamatzky_Costello_2012,Adamatzky_DurandLose_2012} for
reviews).  In their approach, localized excitations (wave fragments) are sent on
ballistic trajectories in an architecture less excitable medium.  Computation is the
result of the interactions between the localized excitations, and complicated logic
circuits can be implemented in this collision-based computing paradigm.


In general, chemistry does not naturally perform these kinds of operations. In other
words, substantial human intervention is required to produce chemical-logic gates and
it is interesting to consider for a moment the forms of these interventions. In a few
rare cases, such as template-molecule based DNA-computing or gene-regulatory
networks, researchers can select the reactants so as to design reaction
networks. This is only possible for rather small reaction networks, with relatively
recent work \cite{Moon_etal_2012} reporting that ``the largest and most complex
program constructed so far'' consists of three gates, built using eleven regulatory
proteins. This pales in comparison to silicon based logic gates (modern CPUs have
hundreds of millions of gates), but evolution works under different constraints and
to accomplish different goals than human engineered computers. It will be interesting
to see how this area advances.

When it is not possible to arbitrarily design reaction networks (and this is the more
typical scenario), the alternative is to construct an environment (\ie `external' or
`boundary' conditions) in which the chemistry operates, so as to cause it to perform
the desired operation. This is most easily seen in Adamatzky's and other's work in
excitable media, where the shape of the spatial boundaries cause the propagating
waves to interact as desired, but the same idea underlies the development of
logic-gates based upon bistable reactions in CSTRs \cite{Lebender_Schneider_1994}, or
light modulated Belousov-Zhabotinsky micro-droplets~\cite{Wang_etal_2016}, etc. In
these latter cases the external conditions are not (only) the shape or properties of
the containers, but also include feed-rates, lighting conditions or other phenomena
that influence the chemical reactions in various ways.


It is well established that varying the external conditions of chemical systems can
change their dynamical regimes. This change can be quantitative, such as when light is
used to influence decay or reaction rates
\cite{Garcia_Sancho_1999,Horsthemke_Lefever_2006,Gagnon_etal_2015,Gagnon_PerezMercader_2017,Gagnon_etal_2017}, or qualitative, when a parameter crosses a bifurcation point, such as the use of
light to suppress oscillations in the chlorine dioxide-iodine-malonic acid reaction~\cite{Munuzuri_etal_1999}.

In all cases that we are aware of, the developers of chemical logic gates have
employed \emph{static} external conditions, \ie conditions that do not change during
the evaluation of logical-operation(s)~\footnote{In some cases (\eg \cite{Wang_etal_2016}) conditions are varied over time so as to provide different \emph{input} to the network. As shall become clear, this is different than the method presented here, where external conditions are regulated \emph{independently} of input, and with the fundamentally different purpose of modifying dynamics into a logical operation.}.  In this paper, we show that in some cases a \emph{sequence of external conditions} is sufficient to transform a well-stirred chemical system (sometimes referred to as the Selkov-Gray-Scott model~\cite{Selkov_1968,Gray_Scott_1985}) into
one capable of logical operations.  Note that the method presented here is conceptually very different from other approaches in the field.  Instead of designing a (sometimes extremely) fine-tuned chemical system with fixed external conditions to perform a logical operation, we subject an existing chemical system to a prescribed temporal sequence of changes in its external conditions.  We argue that this type of reasoning is more likely to be applicable to biological systems. 

The rest of this paper is organized as follows.  Because the method we are presenting is new, we first provide in Sect.~\ref{sec:SPS_toy_model} a didactic example, where our `dynamic external condition' method is applied to a minimal mathematical toy example.  We then show that the method can transform a well-known cubic autocatalytic chemical model into a logic gate in Sect.~\ref{sec:SPS_chemical_model}.  We finally discuss the applicability of the method to biological and other systems in Sect.~\ref{sec:Discussion}.

\section{Dynamic external conditions in a simple toy example \label{sec:SPS_toy_model}}

\begin{figure}[t]
\centering
\includegraphics[width=0.6\linewidth]{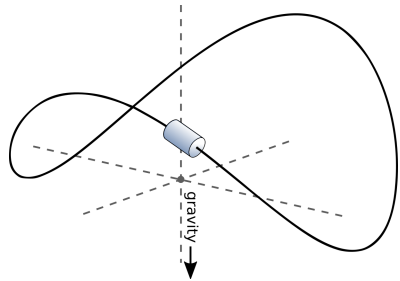}
\caption{Bead moving on a saddle-shaped wire used as a toy model to implement a NOT gate.  Note that the projection of the above saddle-shaped wire on the horizontal plane gives a circle (as depicted in Fig.~\ref{fig:Toy_model_real_space}) and that gravity is pointing down. \label{fig:Double_hill_model}}
\end{figure}

For illustrative purposes, we now show how modulation of a dynamical system's
parameters can transform it into a logical NOT-gate, which transforms \true input into
\false and \false input into \true (the action of a NOT gate is shown in Table~\ref{tab:NOT}).
For our base dynamical system we will consider a bead on a saddle-shaped wire
(Fig.~\ref{fig:Double_hill_model}). Depending on where the bead is placed, it will
slide into one of the two `troughs'.  To identify such a dynamical system with a
logic gate, it is necessary to specify what is meant by inputs and outputs for that
gate.  In this toy example, we decide that inputs are determined by which half-side
of the wire the bead is initially placed and outputs are determined by which trough
the bead ends up in. For both input and output, we consider the right (left) half to
correspond with a \true (\false) value.  The motion of the bead represents the operation of the gate.

\begin{table}[h]
\caption{NOT logic table \label{tab:NOT}}
\centering
\begin{tabular}[c]{c|c}
   Input  & Output \\
   \hline
   \true &  \false \\
   \false &  \true \\
\end{tabular}
\end{table}

As it stands, this system does not represent a NOT-gate. The bead ends in the same trough in which it starts and so the system can be considered an IDENTITY-gate, where \true input produces \true
output and \false input produces \false output.  Indeed no matter the shape of the wire, no \emph{static} configuration of the wire would be able
to accomplish a NOT operation. To cause the bead to switch sides and thus the system
to act as a NOT-gate, we can subject the saddle-shaped wire to a predetermined
sequence of rotations~\footnote{We assume in this toy model that the position of the bead is unchanged by these rotations.}  that take place after the bead is placed in
a way that is entirely independent of the bead's location. The rotations are
comparable to changes in the external conditions of the bead.

\begin{figure}[t]
\centerline{\includegraphics[width=0.55\linewidth]{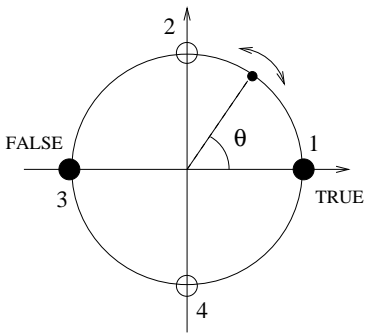}}
\caption{Dynamical system where a bead (small black circle) moves on a circle according to Eq.~\eqref{eq:Simple_toy_model} (with $\phi = 0$).  Large numbered black and white circles represent attractive ($\theta_{1}^{*}$, $\theta_{3}^{*}$) and repulsive ($\theta_{2}^{*}$, $\theta_{4}^{*}$) fixed points, respectively. In our toy example, we assign $\theta_{1}^{*}$ to mean \true and $\theta_{3}^{*}$ to mean \false. \label{fig:Toy_model_real_space}}
\end{figure}

We make these ideas mathematically precise by considering the following
one-dimensional periodic dynamical system:
\begin{eqnarray}
\label{eq:Simple_toy_model}
\frac{d\theta(t)}{dt} & = & \sin \big ( 2\theta(t) - \phi \big),
\end{eqnarray}
where $\theta(t)$ denotes the angular position of a bead moving on a circle and
$\phi$ is a phase parameter.  The angular velocity of the bead depends on its
position and can be either positive (counterclockwise) or negative (clockwise),
depending on the sign of the sine function in Eq.~\eqref{eq:Simple_toy_model}.  The
fixed points of the system are $\theta_{1}^{*} = \frac{\phi}{2}$,
$\theta_{2}^{*} = \frac{\pi}{2} + \frac{\phi}{2}$,
$\theta_{3}^{*} = \pi + \frac{\phi}{2}$ and
$\theta_{4}^{*} = \frac{3\pi}{2} + \frac{\phi}{2}$.  $\theta_{1}^{*}$ and
$\theta_{3}^{*}$ are attractive fixed points, while $\theta_{2}^{*}$ and
$\theta_{4}^{*}$ are repulsive.  The phase, $\phi$, implements rotations of the
saddle-shaped wire around the vertical axis of
Fig.~\ref{fig:Double_hill_model}, as discussed below.   When $\phi = 0$, for any initial condition in the right (left) half-circle $\theta_{\rm RHC}$ ($\theta_{\rm LHC}$), the bead asymptotically goes the to fixed point $\theta_{1}^{*}$ ($\theta_{3}^{*}$). The system thus has two basins of attraction, and the separatrix, \ie the manifold separating those basins, is the points $\theta_{2}^{*}$ and $\theta_{4}^{*}$.  As described above, we associate the attractor of each of these basins with Boolean
values, such that $\theta_{1}^{*} \equiv$ \true and $\theta_{3}^{*} \equiv$ \false.  
Fig.~\ref{fig:Toy_model_real_space} illustrates this dynamical system for $\phi = 0$,
showing the saddle-shaped wire of Fig.~\ref{fig:Double_hill_model} viewed from above,
with the repulsive fixed points of the crests and attractive fixed points of the
troughs indicated as open and closed circles respectively.

As already described informally, for $\phi = 0$, this dynamical system operates as
an IDENTITY-gate, transforming initial conditions associated with a \true input
$\theta(t=0) \in \theta_{\rm RHC}$ into a \true output-state and initial conditions
associated with a \false input $\theta(t=0) \in \theta_{\rm LHC}$, into a \false
output-state. We now show how by judiciously modifying the external conditions, it is
possible to transform this dynamical system into a logical NOT gate.  More
explicitly, for initial conditions in or near the \true state (\ie
$\theta(t=0) \in \theta_{\rm RHC}$) we want the system to approach (at asymptotically
large times) the \false state and inversely, for initial conditions in or near the
\false state (\ie $\theta(t=0) \in \theta_{\rm LHC}$), the system should approach the
\true state.  To accomplish this goal, we first identify a `handle', \ie an `external
condition' ---a parameter that we can modulate.  Examples of such parameters in
chemical systems include flow rates, temperature, stirring rates, etc.  For this
exercise, we use the phase parameter, $\phi$, as an externally adjustable
parameter. We also assume (both here and in the subsequent chemical model) that
changes to external conditions are fast compared to the other dynamics so that
changes in parameter regimes can be approximated as discrete jumps.  When this
assumption does not hold, additional equations describing this parametric dynamics
can be coupled to the main dynamics.

\begin{figure*}[t]
\centerline{\includegraphics[width=0.9\linewidth]{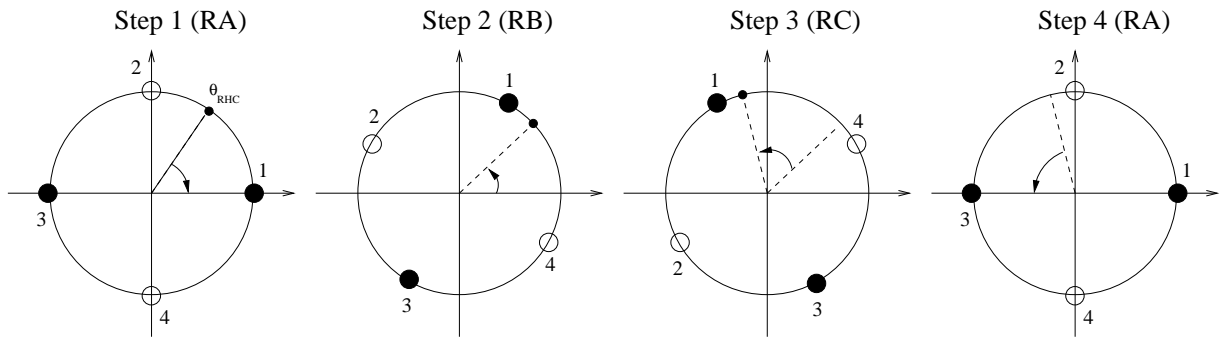}}
\caption{Example of dynamic external conditions necessary to turn the toy dynamical system~\eqref{eq:Simple_toy_model} into a NOT gate.  See text for explanations of the steps.  The labels 'RA', 'RB', 'RC' correspond to the three regimes discussed in the main text. \label{fig:Toy_model_sequential_shifts}}
\end{figure*}

It is now possible to identify a temporal sequence of external conditions that
produces the desired behaviour One such sequence is shown in
Fig.~\ref{fig:Toy_model_sequential_shifts}.  In Step~1, the system is placed in \emph{Regime-A} (RA), where $\phi = \phi_{RA} = 0$. The
system evolves in this regime for the duration $\tau_{RA}$, which is chosen to be
sufficiently long for the bead to approach close to the fixed point centred within
the basin in which it started, \ie
$\theta(t=0) \in \theta_{\rm RHC} \rightarrow \theta_{1}^{*}$ and
$\theta(t=0) \in \theta_{\rm LHC} \rightarrow \theta_{3}^{*}$.  In Step~2, the
system's dynamics are changed to \emph{Regime-B} (RB) by setting the external condition
$\phi = \phi_{RB} = \frac{2\pi}{3}$ for a duration $\tau_{RB}$. This rotates all
fixed points counterclockwise by 60 degrees, and for the duration $\tau_{B}$ the bead will be attracted to the new position of the fixed points whether it started this step in $\theta_{1}^{*}$ or $\theta_{3}^{*}$.  In Step~3, we enter \emph{Regime-C} (RC) by setting $\phi = \phi_{RC} = \frac{4\pi}{3}$
for a time $\tau_{RC}$.  In this parameter regime, the fixed points are rotated
counterclockwise by a further 60 degrees, and again the bead follows.  Finally, in
Step~4, the system is returned to its original configuration, \emph{Regime-A}
($\phi = \phi_{RA} = 0$) for a time $\tau_{4}$.  Since the bead is now on the
opposite side of the separatrix from which it started, it moves to the fixed point
that is on the opposite side of where it began. To summarize, initial conditions
associated with a \true input now result in in a \false output-state
$\theta(t=0) \in \theta_{\rm RHC} \rightarrow \theta(t) = \theta_{3}^*$, and initial
conditions associated with a \false input result in a \true output-state
$\theta(t=0) \in \theta_{\rm LHC} \rightarrow \theta(t) = \theta_{1}^*$. We have thus
transformed the system into a NOT-gate.  

Before we introduce our chemistry-based example, a few further comments are in order.
First, we note that everything about the sequence of the external conditions is
independent of the initial condition of the state of the system. Put another way,
there are no rules that say, for instance, to change to Regime-X \emph{if} the system
is in a particular state or \emph{if} the input is \true, etc. We emphasize this so
as to make clear that the logical operation is performed within the modulated system
---not outside of it, \ie not by the modulation itself.

Second, we observe that the dynamic sequence of external conditions shown in
Fig.~\ref{fig:Toy_model_sequential_shifts} is not unique.  The values of $\phi_{i}$
and $\tau_{i}$ for each step are constrained, but substantial freedom is allowed in
their choice.  For instance, the duration of Step~2 ($\tau_{RB}$) must be long enough
for the bead to move at least 30 degrees counterclockwise, but apart from that is not
constrained.  Similarly, the duration $\tau_{RC}$ must be sufficiently long for the
bead to cross the separatrix that is present in the \emph{Regime-A} state, but is
otherwise unconstrained. Also, other regimes could have been used, for instance,
replacing all $\phi_{i} \rightarrow -\phi_{i}$ would accomplish the same final
operation by moving the fixed points clockwise instead of counterclockwise.

Finally, we observe that this example has treated the simple case of a NOT gate,
which has only one input (associated with the initial condition of the dynamical
system).  Other important gates (AND, OR, NAND, etc) have two inputs.  To be able to
implement those gates using dynamic external conditions, the two inputs must be
combined into one initial condition for the dynamical system.  For instance, a linear
combination of the two inputs could be used.  An explicit example of this in a
chemical context is presented in the next section.

It is important to emphasize that in our method, external conditions are regulated by an `open-loop' controller, where external conditions are regulated according to a pre-determined sequence that is entirely independent of both the current state of the chemistry and the input given to the system. The logical operation is thus performed by the chemical medium and not by the process regulating it.

\section{Dynamic external conditions in a chemical model \label{sec:SPS_chemical_model}}

We now show how a cubic-autocatalytic chemical reaction can be turned into a NAND
gate by dynamic modification of its external conditions.  The action of a NAND gate is shown in Table~\ref{tab:NAND}.  The NAND logic gate was selected for its universality, \ie the fact that all logical operations can be constructed out of networks of this gate~\cite{Sheffer_1913}.
The chemical model (sometimes referred to as the Selkov-Gray-Scott model) was selected as it is a well studied system that is known to be capable of interesting and complicated dynamics~\cite{Gray_Scott_1985}.  It is given by:
\begin{eqnarray}
\label{eq:Gray_Scott_1}
\deriv{U} &= -r_u U - \lambda U V^2 + F_u, \\
\label{eq:Gray_Scott_2}
\deriv{V} &= -r_v V + \lambda U V^2 + F_v,
\end{eqnarray}
where $U = U(t)$ and $V = V(t)$ are chemical concentrations.  These equations can be
thought of as describing a continuous well-stirred flow reaction system, where
autocatalyst V reacts with substrate U, transforming it into more V according to the
reaction U$+$2V $\rightarrow$ 3V. The reactants are fed into the reactor at rates
described by parameters $F_u$ and $F_v$, and they are removed at rates described by
terms $r_u U$ and $r_v V$.  Without loss of generality, we choose the values
$r_{u} = 1.5$, $r_{v} = 3$ and $\lambda = 1$ in the following; other values would change the details
of our results, but not the overall reasoning.

\begin{table}[h]
\caption{NAND logic table \label{tab:NAND}}
\centering
\begin{tabular}[c]{cc|c}
   Input A & Input B & Output \\
   \hline
   \false & \false & \true \\
   \false & \true & \true \\
   \true & \false & \true \\
   \true & \true & \false
\end{tabular}
\end{table}

As was the case for the toy model, a first step is to find parameters that can be
adjusted externally.  We pick the `feed-rate' parameters, $F_u$ and $F_v$, as our
tunable external conditions. In other situations it may make sense to pick other
parameters.  For example if the reaction rate ($\lambda$) were influenced by the
presence of light, $\lambda$ could be used instead of (or in addition to) the
feed-rate parameters.

\begin{figure}[t]
  \centering
  \begin{tabular}{r}
    \includegraphics[width=1.\linewidth]{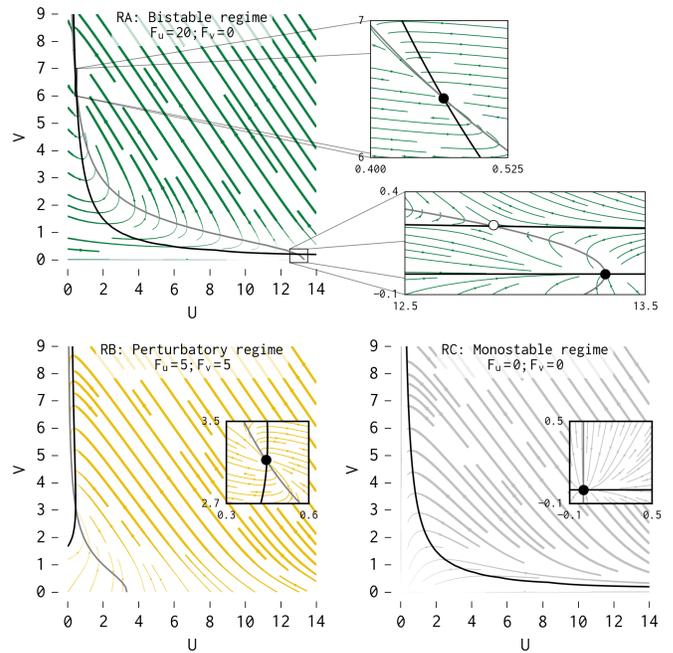}\\ 
  \end{tabular}  \caption{Dynamics for three different parameter regimes of the cubic-autocatalytic chemical reaction represented by Eqs.~\ref{eq:Gray_Scott_1}-\ref{eq:Gray_Scott_2}. $V$ and $U$ nullclines are indicated by black and gray curves respectively. Filled and open circles represent stable and unstable fixed points, respectively. Note that in some insets, non-sensical negative values of $U$ and $V$ are shown to provide insight into the stability of the fixed points. \label{fig:regimes}}
\end{figure}

Once again, there is more than one way to modulate the parameters to cause it to
operate as a logic gate.  It is useful (but not necessary -- see discussion) to have two fixed points that we can associate with the two Boolean values.  We thus start by identifying a configuration of the adjustable parameters for which the system is bistable (see Appendix~\ref{sec:FP} for an analysis of the fixed points of the model). These values are indicated in the row
labelled RA (for Regime-A) in Table~\ref{tab:regime_parameter_values} and plots in
Fig.~\ref{fig:regimes} show the dynamics of the system when the system is in this
configuration. We associate a Boolean value with each of the two fixed points,
arbitrarily selecting $\UVfalse\approx(13.33,0)$ to represent \false and
$\UVtrue\approx(0.5,6.5)$ to represent \true (see Fig.~\ref{fig:regimes}).

\begin{table}[h]
\caption{Parameter regimes \label{tab:regime_parameter_values}}
  \centering
  \begin{tabular}[c]{lcc}
\bfseries Regime & ${\bf F_u}$ & $\bf F_v$ \\
RA (bistable)        & 20 & 0 \\
RB (perturbatory)    & 5  & 5 \\
RC (monostable)      & 0  & 0 \\
\end{tabular}
\end{table}

\begin{figure*}[h]
  \centering
    \includegraphics[width=0.8\linewidth]{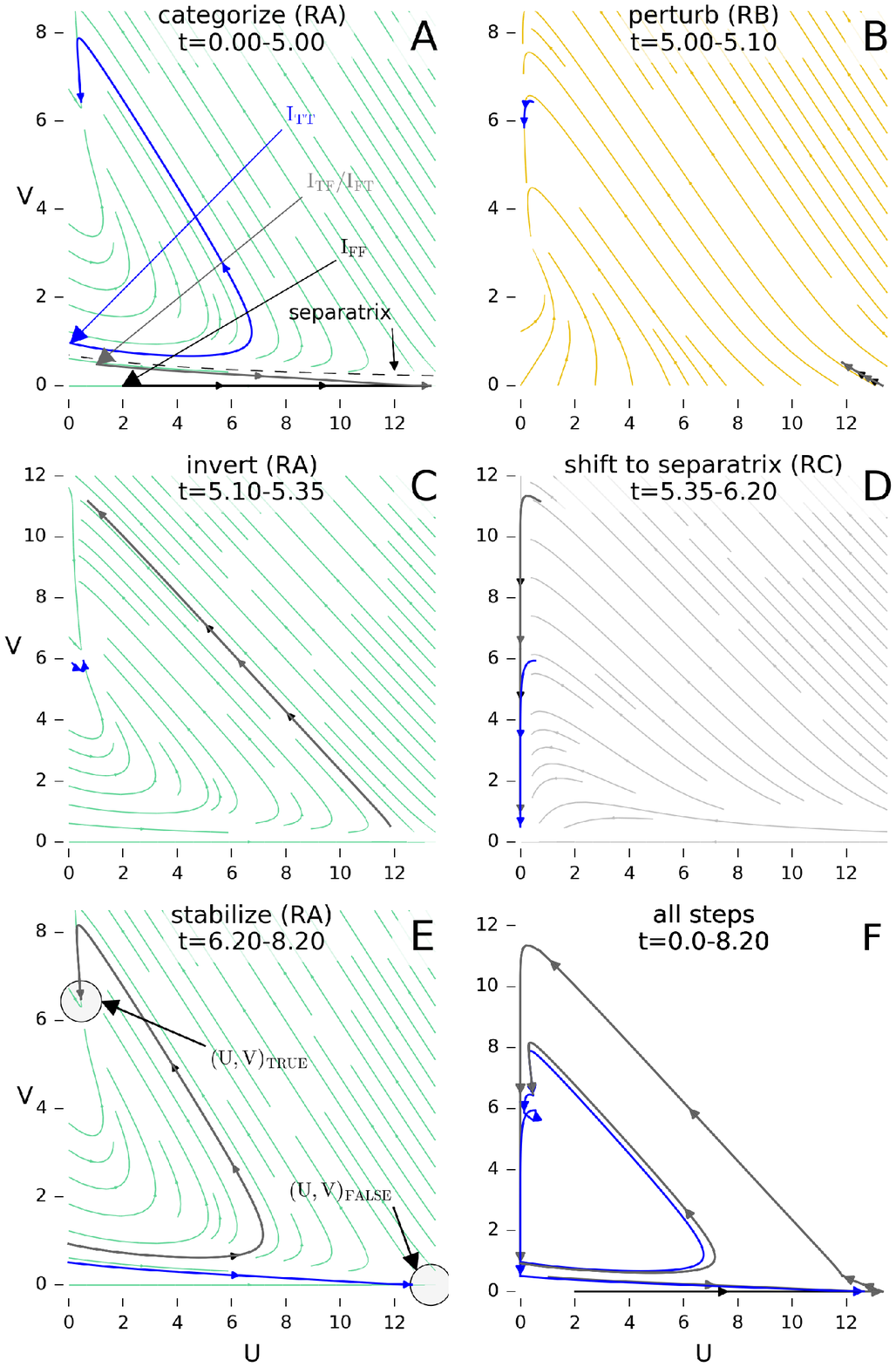}
  \caption{Phase portraits of the cubic-autocatalytic reaction represented by Eqs.~(\ref{eq:Gray_Scott_1})-(\ref{eq:Gray_Scott_2}) performing a NAND operation. Frames A--E each indicate a different phase of the computation where the system is placed in one of three parameter regimes (RA, RB or RC). The duration of each phase is indicated in Table~\ref{tab:regime_sequence} and at the top of each frame. Trajectories from initial conditions corresponding to each of the possible inputs are colour coded such that the blue trajectory corresponds to an input of (\true,\true), the black trajectory corresponds to an input of (\false,\false) and the gray trajectory corresponds to an input of (\true,\false) or (\false,\true). Frame F shows the entire sequence. In Frame A, the separatrix between the two basins of attraction in this regime is indicated by a dashed line.   Note that frames C, D and F have a different vertical scale.   \label{fig:non_spatial} }
\end{figure*}

We assume that the mechanism through which the system receives inputs from upstream
gates is by way of diffusion or some other process, such that the input is the
average of two of the Boolean states, attenuated by some constant, $k$. For example,
if the chemical NAND gate were to be implemented with the use of a CSTR, the two
inputs might correspond to samples taken from upstream CSTR-based gates. The net
effect on the concentration of the reference chemical in the CSTR would thus depend
on the average of the two input concentrations, diluted in the whole reaction volume.
In this example, $k$ represents the effect of dilution.  For the Selkov-Gray-Scott model,
the possible inputs (initial conditions) are thus $I_{FF} = k\UVfalse$,
$I_{TT} = k\UVtrue$, and $I_{TF}=I_{FT}=k(\UVtrue+\UVfalse)/2 = k(6.6,3.25)$, where
$k$ is set to $0.15$.  Similar to system parameters $r_u$,$r_v$ and $\lambda$, the exact value of $k$ is not important for our method to work, but is important in the modeling of realistic physical/chemical situations.  A different value of $k$ might affect the specifics of each regime (timing, values of the parameters), but not the overall recipe.  For instance, in the present case the value of $k$ was chosen such that of the four possible input values, only $I_{TT}$ lies above Regime-A's separatrix (see Frame A in Fig.~\ref{fig:non_spatial}). We confirmed in numerical simulations that small changes to $k$ (we checked $k \in[0.11,..0.18]$) would have no effect upon the logical operation performed. Larger changes to $k$ can change the type of logical operation accomplished. For instance, a $k$ value of $0.25$ places both $I_{TF}$ and $I_{FT}$ above the separatrix, which causes the regime sequence to accomplish a NOR operation instead of NAND (confirmed in numerical simulation, results not shown). Changing regime parameters, or the sequence or duration of regimes used would allow further flexibility in selecting a value for $k$. For instance, the parameter $F_u$ influences the position of the separatrix in Regime-A.

To operate as a NAND gate, trajectories that start at initial conditions $I_{FF}$,
$I_{TF}$, and $I_{FT}$ must end at $\UVtrue$ and initial condition $I_{TT}$ must end
at $\UVfalse$.  There is no single fixed parametric regime of the system that
accomplishes this behaviour. But we now explain how this behaviour can be achieved
when the system is placed in the temporal sequence of external conditions specified
in Table \ref{tab:regime_sequence}.

We start with the system in the Bistable Regime (Regime RA in
Table~\ref{tab:regime_parameter_values}). This categorizes the trajectories such that
those initial conditions that are associated with an output of \true are at one of
the fixed points, while the other initial condition falls into the other fixed point
(see Fig.~\ref{fig:non_spatial}A). If we were implementing an AND gate rather than a
NAND gate, we could stop here, as inputs $I_{FF}$, $I_{TF}$, and $I_{FT}$ are at
\UVfalse and $I_{TT}$ is at \UVtrue. However, AND gates are not universal, and to
accomplish our goal of creating a NAND gate, we must `invert' the system such that
those trajectories at \UVfalse are moved to \UVtrue and vice versa.

\begin{table}[h]
\caption{NAND regime sequence\label{tab:regime_sequence}}
  \centering
  \begin{tabular}[c]{llc}
\bfseries Step Name & \bfseries Regime & \bfseries Duration \\
Categorize & RA (bistable) & 5.00 \\
Perturb & RB (perturbatory) & 0.10 \\
Invert & RA (bistable) & 0.25 \\
Shift to separatrix & RC (monostable) & 0.85 \\
Stabilize & RA (bistable) & 2.00 \\
\end{tabular}
\end{table}

We briefly place the system in the Perturbatory Regime (regime RB in
Table~\ref{tab:regime_parameter_values}), which causes a small increase in $V$ for
all of the trajectories (Step 2, Fig.~\ref{fig:non_spatial}B). This small increase in
$V$ means that when the system is returned to the Bistable Regime in Step 3
(Fig.~\ref{fig:non_spatial}C), the $I_{FF}$, $I_{TF}$, and $I_{FT}$ trajectories are
no longer in the \UVfalse basin of attraction, and they move on a transient that
takes them higher than the $I_{TT}$ trajectory. If we were to stay in the Bistable
Regime for a long time, all of the trajectories would approach \UVtrue. This would be
bad as it would no longer be possible to distinguish between the four inputs. To
prevent this from happening, we move to the Monostable Regime (Step 4,
Fig.~\ref{fig:non_spatial}D) before the system comes to equilibrium. In this regime,
the feed rates $F_u$ and $F_v$ are both $0$, causing all of the trajectories to
decrease in both $U$ and $V$. We again make use of the transient dynamics by moving
the system into the bistable regime before the system comes to equilibrium. With the
correct timing, at the start of the final bistable regime (Step 5,
Fig.~\ref{fig:non_spatial}E) the $I_{FF}$, $I_{TF}$, and $I_{FT}$ are on one side of
the separatrix, within the basin of attraction of \UVtrue and $I_{TT}$ is on the
other side, within the basin of attraction of \UVfalse. Left in this final regime of
the sequence, the $I_{FF}$, $I_{TF}$, and $I_{FT}$ trajectories approach \UVtrue and
the $I_{TT}$ approaches \UVfalse, and we have successfully completed a NAND
operation.

This process involves the sequential use of three different parameter regimes, the
bistable regime (RA), the perturbatory regime (RB), and the monostable regime
(RC). Although the timing of the duration of each regime must be somewhat precise, in
our experience the values of the parameters for the regime do not need to be
precisely tuned, provided they perform the desired qualitative dynamics (\eg
bistability, or desired transient dynamics). The values of the parameters for each
regime can be found in Table \ref{tab:regime_parameter_values} and the order and
duration of regimes for performing a NAND operation in a non-spatial system can be
seen both in Fig.~\ref{fig:non_spatial} and Table \ref{tab:regime_sequence}.


\section{Discussion \label{sec:Discussion}}

The toy and chemical models presented above are two simple examples of dynamical
systems that can be turned into logic gates using dynamic external conditions.  The
method presented here is general and could in principle be applied to other dynamical
systems, including chemical systems.

The NAND-gate that we have implemented includes two properties that are necessary if
the gates are to be assembled into networks of gates that are capable of useful
computation above and beyond a simple NAND operation. The first property is error
correction. Both of the presented systems employ a bistable regime, where two
distinct stable fixed points are used to correct for noise or small perturbations in the initial conditions, and steering the trajectory to a well-defined, time-invariant state that is a function of the
input. This error-correcting `categorization' step limits the influence of
initial conditions upon subsequent steps. Without this error correction, noise can
accumulate within sequences of gates, making them ineffective when combined into
larger computational circuits thus limiting the possibility of composing them into
useful computational networks. Ref.~\cite{Magnasco_1997} discusses this idea in the
context of chemical neurons.

The second property concerns the transmission of the output of one logic-gate into
the input of a subsequent `downstream' gate. If chemical logic gates are to be
composed into networks capable of more complex computation, it must be possible to
use the output of certain gates as the input of others. Not all previous research on
chemistry-based logic gates has fully addressed this demanding requirement. For
example, in Ref.~\cite{Wang_etal_2016}, input to a chemical logic gate is
defined by the presence or absence of oxidation spikes in the BZ reaction of two
`input' droplets, whereas the output in this system is defined by \emph{a delay} in
the autonomous oscillations of a third `output' droplet. Without further work it is
unclear how this output could be used as the input for a subsequent downstream
gate. The idea here that we want to emphasize is that for chemical logic gates to be
composable into networks, the `format' or `representation' of output Boolean values
must be the same as (or compatible with) that of the input. Our method allowed us to
produce a chemical logic gate where the output of the gate, once diffused and
diluted, could be used as input for another gate.

In general, the process of designing of a specific gate using dynamic external
conditions depends strongly on the underlying system and on the number and effect(s)
of the externally adjustable parameters.  In some cases there will be no solution; no
sequence of external conditions will cause the base system to perform the desired
behaviour. In general, it is difficult to establish whether or not a solution exists
but occasionally, when certain fundamental properties are absent from the base system
or when the effects of external conditions are limited, the absence of a solution can
be quickly established. For instance if a system always has a single basin of
attraction, regardless of any external parameter shifts, it seems that it would be
impossible to create a NAND-gate with error correction. Our example systems employ
two stable fixed-points to accomplish error correction, but it is interesting to
speculate that, one could replace one or more fixed points with limit cycles or
chaotic attractors ---provided those attractors are sufficiently separated in phase
space, then they could be used as Boolean value representations.  Multistability
could in principle be used to implement n-valued logic gates and that other forms of
signal-processing, such as amplification, dynamic range compression, etc. could be
accomplished with monostable or other systems.

For the presented method to work, the tunable external conditions must have
sufficient influence upon the system's dynamics. Typical external conditions for
chemistry includes flow rate, temperature, stirring rate, and light sensitive
reaction rate, but in many specific cases these parameters may not sufficiently
influence the chemical dynamics, making it impossible to use the presented method.
Additionally, we have assumed that the parametric shifts can occur quickly compared
to the other dynamics. In some systems it may be possible to relax this assumption,
but in others it may be impossible to regulate the external conditions quickly enough
to avoid undesired transients. Ultimately, the properties of the chemical medium (reaction rates, diffusion rate, etc) in which the computation is taking place imposes a number of unavoidable constraints that might limit the experimental applicability of the method.  For instance, the shift to the separatrix (see Table~\ref{tab:regime_sequence}) might be hard to achieve experimentally if reactions are too fast.

The necessary or desired characteristics for a base system to be amenable
to this method is difficult to describe at this stage.  The method requires a case by
case study of its dynamics in different regimes and trial and error and
experimentation.  From a practical point of view, it means that a mathematical model
of the dynamics is very helpful (perhaps even necessary) to find out all the regimes
and timings for the method to work.  Since models of chemical dynamics are often
approximate, this can add a layer of difficulty when applying this method to complex
chemical systems.

With regards to the chemical processing of information in biological systems, our approach to chemical logic gates is conceptually different from previous studies on the subject.  For instance, the chemical logic gates of Refs.~\cite{Hjelmfelt_etal_1991,Hjelmfelt_etal_1992a,Hjelmfelt_etal_1992b} require building a precise network of chemical neurons (each made of many chemicals) in order to perform a certain logical operation.  The networks are built specifically for the task they are designed for, and the absence of one or more chemicals could disrupt its proper functioning.  This high level of fine tuning seems to be generic of other chemical gate designs.  This may be hard to achieve naturally in a biological setting.

It is interesting to return for a moment to one of the motivations of natural computing:
understanding how organisms process information.  The approach presented here is
different than previous approaches to chemical logic in that it uses existing
`chemical hardware' (already designed for certain tasks) to process information.  By
playing on the external conditions of some pre-existing chemical dynamics, cells may
be able to perform a logical operation by re-purposing some of its internal chemical
components by dynamically regulating the external conditions of that chemistry (\eg
transport of reactants through the cell membrane).  In addition, our two examples
explicitly show that there is no unique temporal sequence of external conditions and
also substantial freedom in the choice of parameters in our method.  So it is not
unreasonable to think that evolution could tweak relatively easily the external
conditions of biological processes in order for them to perform logical operations or
other forms of signal manipulation.  This makes the above method a potentially useful
way of processing information in biological systems. We are not aware of any example of natural or synthetic biological systems that uses such a method, but it would be interesting to pursue research in that direction.

\section{Conclusion \label{sec:Conclusion}}

We have shown that a well-known cubic autocatalytic chemical model can be transformed
into a logic gate by changing its external conditions according to a prescribed
time-dependent sequence that is independent of the state of the system and of the
input to the system.  The method is general and could in principle be applied to
other dynamical systems. In a parallel publication we present our application of the
method to a spatial model (where diffusion effects are important), showing how the
method can be used to take advantage of spatial symmetries and produce a network of
logic gates out of a uniform chemical medium.


\appendix

\section{Fixed points of the Selkov-Gray-Scott model\label{sec:FP}}

 The fixed points of the Selkov-Gray-Scott model in the case where $F_{v}=0$ are obtained by setting Eqs.~(\ref{eq:Gray_Scott_1})-(\ref{eq:Gray_Scott_2}) to zero.  There are three fixed points in total:
\begin{eqnarray}
\label{eq:FP1}
\left(U_{1}^{*},V_{1}^{*}\right) & = & \left(\frac{F_{u}}{r_{u}}, 0\right) \\
\label{eq:FP23}
\left(U_{2,3}^{*},V_{2,3}^{*}\right) & = & \left(\frac{F_{u}\pm\sqrt{F_{u}^{2}-\frac{4r_{u}r_{v}^{2}}{\lambda}}}{2r_{u}}, \frac{F_{u}\mp\sqrt{F_{u}^{2}-\frac{4r_{u}r_{v}^{2}}{\lambda}}}{2r_{v}}\right) \nonumber \\
\end{eqnarray}
The stability of the system can be obtained by linearising Eqs.~(\ref{eq:Gray_Scott_1})-(\ref{eq:Gray_Scott_2}) around each fixed point~\cite{Strogatz_1994}.  The linearised system is:
\begin{equation}
\left(\begin{array}{c} \frac{d(\delta U)}{dt} \\[3pt] \frac{d(\delta V)}{dt} \end{array}\right) = \left(\begin{array}{cc} -r_{u} -\lambda V^{2} & -2\lambda UV \\ \lambda V^{2} & -r_{v} + 2\lambda UV \end{array}\right)_{U^{*},V^{*}} \left(\begin{array}{c} \delta U \\ \delta V\end{array}\right)
\end{equation}
where $\delta U$, $\delta V$ represent small deviations in concentrations around the fixed point $(U^{*},V^{*})$.  For the parameter values specified below Eqs.~(\ref{eq:Gray_Scott_1})-(\ref{eq:Gray_Scott_2}) and in Table~\ref{tab:regime_parameter_values}, we get that regime-A has two stable fixed points at $(13.33,0)$ and $(0.46,6.43)$ and one unstable fixed point at $(12.87,0.23)$, while regime-C has only one stable fixed point at $(0,0)$.  The analysis of regime-B is more complicated, since the expressions for the fixed points~(\ref{eq:FP1})-(\ref{eq:FP23}) are obtained under the assumption that $F_{v}=0$.  Analytical expressions for the fixed points in regime-B exist, but they are not particularly illuminating and we do not include them here.  Nonetheless, it is possible to show that for the parameter values specified in Table~\ref{tab:regime_parameter_values}, there is a single stable fixed point at $(0.45,3.11)$.

\section*{Competing interests \label{sec:Competing_interests}}

The authors have no competing interests.

\section*{Authors' contribution \label{sec:Contribution}}

ME conceived of the study, carried out the numerical simulations, and participated in the writing of the manuscript; JSG conceived of the study, contributed in the more theoretical aspects of the study, and participated in the writing of the manuscript; JPM coordinated the study, participated in its design, and helped draft the manuscript.  All authors gave final approval for publication.

\begin{acknowledgments}
The authors thank J. Szymanski for useful discussions. 
\end{acknowledgments}

\section*{Funding \label{sec:Funding}}

This research is supported by Repsol S.A.

\bibliographystyle{unsrt}
\bibliography{chemical_logic_gate_nonspatial}

\begin{thebibliography}{10}

\bibitem{Rozenberg_etal_2012}
G.~Rozenberg, T.~B\"{a}ck, and J.~N. Kok.
\newblock {\em Handbook of Natural Computing}.
\newblock Springer, 2012.

\bibitem{Bray_1995}
D.~Bray.
\newblock Protein molecules as computational elements in living cells.
\newblock {\em Nature}, 376:307, 1995.

\bibitem{Mitchell_2011}
M.~Mitchell.
\newblock Biological computation.
\newblock {\em Ubiquity}, page~1, 2011.

\bibitem{Beer_1997}
R.~Beer.
\newblock The dynamics of adaptive behavior: A research program.
\newblock {\em Robotics and Autonomous Systems}, 20:257, 1997.

\bibitem{van_Gelder_1998}
T.~van Gelder.
\newblock The dynamical hypothesis in cognitive science.
\newblock {\em Behavioral and Brain Sciences}, 21:615, 1998.

\bibitem{Rossler_1974}
O.~R\"{o}ssler.
\newblock In M.~Conrad, W.~G\"{u}ttinger, and M.~Dal~Cin, editors, {\em Lecture
  Notes in Biomathematics 4}, page 546. Springer, 1974.

\bibitem{Okamoto_etal_1987}
M.~Okamoto, T.~Sakai, and K.~Hayashi.
\newblock Switching mechanism of a cyclic enzyme: Role as a ``chemical diode''.
\newblock {\em BioSystems}, 21:1, 1987.

\bibitem{Okamoto_1992}
M.~Okamoto.
\newblock Biochemical switching device: biomimetic approach and application to
  neural network study.
\newblock {\em Journal of Biotechnology}, 109:109, 1992.

\bibitem{Hasty_etal_2002}
J.~Hasty, D.~McMillen, and J.~J. Collins.
\newblock {Engineered gene circuits}.
\newblock {\em Nature}, 420:224, 2002.

\bibitem{Wang_etal_2011}
B.~Wang, R.~I. Kitney, N.~Joly, and M.~Buck.
\newblock Engineering modular and orthogonal genetic logic gates for robust
  digital-like synthetic biology.
\newblock {\em Nature Communications}, 2:508, 2011.

\bibitem{Bradley_etal_2016}
R.~W. Bradley, M.~Buck, and B.~Wang.
\newblock Recognizing and engineering digital-like logic gates and switches in
  gene regulatory networks.
\newblock {\em Current Opinion in Microbiology}, 33:74, 2016.

\bibitem{Ashkenasy_etal_2011}
G~Ashkenasy, Z.~Dadon, S.~Alesebi, N.~Wagner, and N.~Ashkenasy.
\newblock Building logic into peptide networks: Bottom-up and top-down.
\newblock {\em Isr. J. Chem.}, 51:106, 2011.

\bibitem{Paun_etal_2005}
G.~Paun, G.~Rozenberg, and A.~Salomaa.
\newblock {\em {DNA computing: New computing paradigms}}.
\newblock Springer, 2005.

\bibitem{Wang_etal_2016}
A.~L. Wang, J.~M. Gold, N.~Tompkins, M.~Heymann, K.~I. Harrington, and
  S.~Fraden.
\newblock Configurable nor gate arrays from belousov-zhabotinsky
  micro-droplets.
\newblock {\em Eur. Phys. J. Special Topics}, 225:211, 2016.

\bibitem{Laplante_etal_1995}
J.~P. Laplante, M.~Pemberton, A.~Hjelmfelt, and J.~Ross.
\newblock Experiments on pattern recognition by chemical kinetics.
\newblock {\em J. Phys. Chem.}, 99:10063, 1995.

\bibitem{DeSilva_Uchiyama_2007}
A.~P. De~Silva and S.~Uchiyama.
\newblock Molecular logic and computing.
\newblock {\em Nature nanotechnology}, 2:399, 2007.

\bibitem{Dittrich_etal_2001}
P.~Dittrich, J.~Ziegler, and W.~Banzhaf.
\newblock Artificial chemistries -- a review.
\newblock {\em Artificial Life}, 7:225, 2001.

\bibitem{Hjelmfelt_etal_1991}
A.~Hjelmfelt, E.~D. Weinberger, and J.~Ross.
\newblock {Chemical implementation of neural networks and Turing machines}.
\newblock {\em Proc. Nati. Acad. Sci. USA}, 88:10983, 1991.

\bibitem{Hjelmfelt_etal_1992a}
A.~Hjelmfelt, E.~D. Weinberger, and J.~Ross.
\newblock {Chemical implementation of finite-state machines}.
\newblock {\em Proc. Nati. Acad. Sci. USA}, 89:383, 1992.

\bibitem{Hjelmfelt_etal_1992b}
A.~Hjelmfelt and J.~Ross.
\newblock {Chemical implementation and thermodynamics of collective neural
  networks}.
\newblock {\em Proc. Nati. Acad. Sci. USA}, 89:388, 1992.

\bibitem{Toth_Showalter_1995}
\'{A}. T\'{o}th and K.~Showalter.
\newblock Logic gates in excitable media.
\newblock {\em J. Chem. Phys.}, 103:2058, 1995.

\bibitem{Steinbock_etal_1996}
O.~Steinbock, P.~Kettunen, and K.~Showalter.
\newblock Chemical wave logic gates.
\newblock {\em J. Phys. Chem.}, 100:18970, 1996.

\bibitem{Adamatzky_2004}
A.~Adamatzky.
\newblock Collision-based computing in belousov-zhabotinsky medium.
\newblock {\em Chaos, Solitons and Fractals}, 21:1259, 2004.

\bibitem{Costello_Adamatzky_2005}
B.~De~Lacy~Costello and A.~Adamatzky.
\newblock Experimental implementation of collision-based gates in
  belousov-zhabotinsky medium.
\newblock {\em Chaos, Solitons and Fractals}, 25:535, 2005.

\bibitem{Adamatzky_etal_2005}
A.~Adamatzky.
\newblock {\em Reaction-diffusion computers}.
\newblock Elsevier, 2005.

\bibitem{Adamatzky_Costello_2012}
A.~Adamatzky and B.~De~Lacy~Costello.
\newblock Reaction-diffusion computing.
\newblock In G.~Rozenberg, T.~B\"{a}ck, and J.~N. Kok, editors, {\em Handbook
  of Natural Computing}, volume~4, page 1898. Springer, 2012.

\bibitem{Adamatzky_DurandLose_2012}
A.~Adamatzky and J.~Durand-Lose.
\newblock Collision-based computing.
\newblock In G.~Rozenberg, T.~B\"{a}ck, and J.~N. Kok, editors, {\em Handbook
  of Natural Computing}, volume~4, page 1949. Springer, 2012.

\bibitem{Moon_etal_2012}
T.~S. Moon, C.~Lou, A.~Tamsir, B.~C. Stanton, and C.~A. Voigt.
\newblock Genetic programs constructed from layered logic gates in single
  cells.
\newblock {\em Nature}, 491:249, 2012.

\bibitem{Lebender_Schneider_1994}
D.~Lebender and F.~W. Schneider.
\newblock Logical gates using a nonlinear chemical reaction.
\newblock {\em J . Phys. Chem.}, 98:1533, 1994.

\bibitem{Garcia_Sancho_1999}
J.~Garcia-Ojalvo and J.~M. Sancho.
\newblock {\em Noise in Spatially Extended Systems}.
\newblock Springer, 1999.

\bibitem{Horsthemke_Lefever_2006}
W.~Horsthemke and R.~Lefever.
\newblock {\em Noise-induced transitions}.
\newblock Springer, 2006.

\bibitem{Gagnon_etal_2015}
J.~S. Gagnon, D.~Hochberg, and J.~P{\'e}rez-Mercader.
\newblock Small-scale properties of a stochastic cubic-autocatalytic
  reaction-diffusion model.
\newblock {\em Phys. Rev. E}, 92:042114, 2015.

\bibitem{Gagnon_PerezMercader_2017}
J.~S. Gagnon and J.~P{\'e}rez-Mercader.
\newblock {Clues on chemical mechanisms from renormalizability: The example of
  a noisy cubic autocatalytic model}.
\newblock {\em Physica}, A480:51--62, 2017.

\bibitem{Gagnon_etal_2017}
J.~S. Gagnon, D.~Hochberg, and J.~P\'erez-Mercader.
\newblock Effects of spatial and temporal noise on a cubic-autocatalytic
  reaction-diffusion model.
\newblock {\em Phys. Rev. E}, 95:032106, Mar 2017.

\bibitem{Munuzuri_etal_1999}
A.~P. Mu\~{n}uzuri, M.~Dolnik, A.~M. Zhabotinsky, and I.~R. Epstein.
\newblock Control of the chlorine dioxide-iodine-malonic acid oscillating
  reaction by illumination.
\newblock {\em J. Am. Chem. Soc.}, 121:8065--8069, 1999.

\bibitem{Note1}
In some cases (e.\protect \tmspace +\thinmuskip {.1667em}g.\protect \xspace
  \cite {Wang_etal_2016}) conditions are varied over time so as to provide
  different \protect \emph {input} to the network. As shall become clear, this
  is different than the method presented here, where external conditions are
  regulated \protect \emph {independently} of input, and with the fundamentally
  different purpose of modifying dynamics into a logical operation.

\bibitem{Selkov_1968}
E.~E. Sel'kov.
\newblock Self-oscillations in glycolysis.
\newblock {\em European J. Biochem.}, 4:79, 1968.

\bibitem{Gray_Scott_1985}
P.~Gray and S.~K. Scott.
\newblock {Sustained oscillations and other exotic patterns of behavior in
  isothermal reactions}.
\newblock {\em J.Phys.Chem.}, 89:22--32, 1985.

\bibitem{Note2}
We assume in this toy model that the position of the bead is unchanged by these
  rotations.

\bibitem{Sheffer_1913}
H.~M. Sheffer.
\newblock A set of five independent postulates for boolean algebras, with
  application to logical constants.
\newblock {\em Transactions of the American Mathematical Society}, 14:481,
  1913.

\bibitem{Magnasco_1997}
M.~O. Magnasco.
\newblock Chemical kinetics is turing universal.
\newblock {\em Phys. Rev. Lett.}, 78:1190, 1997.

\bibitem{Strogatz_1994}
S.~H. Strogatz.
\newblock {\em Nonlinear Dynamics and Chaos}.
\newblock Westview Press, 1994.

\end{thebibliography}

\end{document}